\documentclass[journal]{IEEEtran}
\usepackage{amsmath,amsfonts}
\usepackage{algorithmic}
\usepackage{algorithm}
\usepackage{array}
\usepackage{textcomp}
\usepackage{stfloats}
\usepackage{url}
\usepackage{verbatim}
\usepackage{graphicx}
\usepackage{cite}
\usepackage{bm}
\usepackage{color}
\usepackage{xcolor}
\usepackage{amsmath,amssymb,amsfonts}
\usepackage{booktabs}
\usepackage{diagbox}

\begin{document}
	
	\title{Generalized Tensor-Aided Channel Estimation for Hardware Impaired Device Identification}
	
	\author{Qi Wu,~\IEEEmembership{Student Member,~IEEE,} Zeping Sui,~\IEEEmembership{Member,~IEEE,} Hien Quoc Ngo,~\IEEEmembership{Fellow,~IEEE,} Qun Wan,~\IEEEmembership{Senior Member,~IEEE,} and Michail Matthaiou,~\IEEEmembership{Fellow,~IEEE}
		\thanks{This work was supported in part by the Project REASON, an U.K. Government Funded Project under the Future Open Networks Research Challenge (FONRC) Sponsored by the Department of Science Innovation and Technology (DSIT), and in part by the U.K. Engineering and Physical Sciences Research Council (EPSRC) under Grant EP/X04047X/1. The work of M. Matthaiou was supported by the European Research Council (ERC) under the European Union's Horizon 2020 Research and Innovation Programme under Grant 101001331. The work of H. Q. Ngo was supported by the U.K. Research and Innovation Future Leaders Fellowships under Grant MR/X010635/1, and a research grant from the Department for the Economy Northern Ireland under the US-Ireland R\&D Partnership Programme. (\emph{Corresponding author: Zeping Sui})}
		\thanks{Q. Wu and Q. Wan are with the School of Information and Communication Engineering, University of Electronic Science and Technology of China, Chengdu 611731, China, and also with the National Key Laboratory on Blind Signal Processing, Chengdu 610041, China (e-mail: wuqi57@std.uestc.edu.cn; wanqun@uestc.edu.cn).}
		\thanks{Z. Sui was with the Centre for Wireless Innovation (CWI), Queen's University Belfast, Belfast BT3 9DT, U.K. He is now with the School of Computer Science and Electronics Engineering, University of Essex, Colchester CO4 3SQ, U.K. (e-mail: zepingsui@outlook.com).}
		\thanks{H. Q. Ngo and M. Matthaiou are with the Centre for Wireless Innovation (CWI), Queen's University Belfast, Belfast BT3 9DT, U.K., and with the Department of Electronic Engineering, Kyung Hee University, Yongin-si, Gyeonggi-do 17104, Republic of Korea (e-mail: \{hien.ngo, m.matthaiou\}@qub.ac.uk).}%
		\vspace{-3em}
	}
	
	
	
	\maketitle
	
	\begin{abstract}
		In this paper, we investigate the joint generalized channel estimation and device identification problem in Internet of Things (IoT) networks {under multipath propagation}. To fully utilize the received signal, we decompose the generalized channel into three components: transmitter hardware characteristics, path gains, and angles of arrival. By modeling the received signals as parallel factor (PARAFAC) tensors, we develop alternating least squares (ALS)-based algorithms to simultaneously estimate the generalized channels and identify the transmitters. Simulation results show that the proposed scheme outperforms {both Khatri-Rao Factorization (KRF) and the conventional least squares (LS) method} in terms of channel estimation accuracy and achieves performance close to the derived Cramér–Rao lower bound.
	\end{abstract}
	
	\vspace{-1em}
	\section{Introduction}
	
	Due to the widespread application of 5G technology, the number of Internet of Things (IoT) devices is increasing rapidly, which entails severe security vulnerabilities. Device identification is essential to achieving communication security by authorizing legitimate users and rejecting malicious ones. Conventional cryptography-based authentication schemes, such as password-based \cite{intro_pass1,intro_pass2} and certificate-based authentication \cite{intro_cer1}, require high computation consumption and storage resources. Therefore, designing a low-cost identification scheme for IoT devices is vital.
	
	Radio frequency fingerprint (RFF)-based device identification has emerged as a promising solution for IoT authentication \cite{intro_rff1,intro_rff2}. In wireless communication systems, imperfections in transmitter hardware components including filters, mixers, oscillators, and power amplifiers leave characteristic imprints on the baseband signal, collectively called RFF \cite{sui2024performance,sui2024star}. By modeling these hardware-induced features as part of the communication channel and integrating them with conventional channel models, a generalized channel can be created for IoT device identification. {This approach provides a hardware-based alternative to cryptographic methods, making it especially suitable for resource-constrained IoT environments. Moreover, it offers enhanced security by exploiting device-specific characteristics that are inherently difficult to replicate, thus mitigating spoofing attacks.} Nevertheless, to the best of our knowledge, {no existing research  has explicitly explored the integration of RFF-based modeling into the IoT device identification space.}

	In general, cascaded radio channels contain three components, i.e., the transmitter, the air and the receiver sides. We note that tensor-based methods have been applied to solve channel estimation with similar structures by utilizing the inherent multidimensional structure of signals and channels \cite{eg2,intro_ceten1,intro_ceten2}.
	The main principle of this approach consists of decomposing the high dimensional tensor data into multiple lower-rank ones and providing an efficient approach for channel estimation, among which the most commonly used is the parallel factor (PARAFAC) model \cite{harshman_PARAFAC,koldatensor}. 
	
	In this paper, based on PARAFAC decomposition, we leverage the regularized alternating least square (ALS) algorithm to perform generalized channel estimation \cite{tensor_als}. The proposed algorithms utilize the unfolded forms of the involved matrices to achieve efficient generalized channel estimation. Simulation results demonstrate that our method provides superior performance {over Khatri-Rao Factorization (KRF) and conventional least squares (LS)}. Our main contributions are summarized as follows:
	\begin{itemize}
		\item \textit{Generalized Channel Model:} We introduce a generalized channel model incorporating three critical components—signal generation with hardware imperfections, transmission with channel fading, and multi-antenna reception.
		
		\item \textit{Tensor-Based Device Identification:} We formulate the channel estimation problem for device identification as a third-order tensor modeling task conforming to the PARAFAC structure. By exploiting the Khatri-Rao structure of the combined channel matrix, we propose a PARAFAC-based ALS (TALS) method and show that it outperforms conventional approaches.
		
		\item \textit{Performance Analysis and CRLB:} We provide a theoretical performance analysis of the proposed TALS algorithm and derive the corresponding Cramér–Rao lower bound (CRLB). Simulation results under diverse scenarios validate the tightness of the derived bound.
	\end{itemize}
	
	
	Notation: We use lower-case bold letters, upper-case bold letters, and calligraphic letters to denote vectors, matrices and tensors, respectively; $\bm a(m)$ represents the $m$-th element of vector $\bm a$, while $\bm A(m,n)$, $\bm A_{m,:}$ and $\bm A_{:,n}$ stand for the $(m,n)$-th element, $m$-th row and $n$-th column of matrix $\bm A$, respectively. A vector of all ones with length $k$ is denoted as $\bm 1_k$, and an identity U-way tensor of dimension $D\times\dots\times D$ is denoted as $\mathcal{I}_{U,D}$. The superscript $(\cdot)^T$, $(\cdot)^H$, $(\cdot)^*$ and $(\cdot)^{-1}$ are the transpose, conjugate transpose, conjugate and inverse operators, respectively; $\left\| \cdot \right\|_F$ and $\left|\cdot\right|$ denote the Frobenius norm and modulus operators, respectively. In addition, $\text{diag}(\cdot)$ and $\text{blkdiag}(\cdot)$ create a diagonal matrix and a block diagonal matrix, respectively; $C^k_n = \frac{n!}{(n-k)!k!}$ returns the binomial coefficient and $n!$ is the factorial of $n$. The symbol $\mathrm{vec}(\cdot)$ represents the operation for converting a matrix into a vector by column-stacking. Finally, the Kronecker product, Khatri-Rao product, $n$-mode product, expectation and real part operators are denoted by $\otimes$, $\odot $, $\times_n$, $\mathbb{E}[\cdot ]$ and $\Re$, respectively. 
	\section{System Model}
	{In IoT networks, device identification is critical for countering spoofing attacks. This work exploits RFFs from transmitter hardware imperfections for device-specific identification without cryptographic methods. We consider an IoT scenario where a multi-antenna access point (AP) acts as the receiver, distinguishing multiple devices based on their RFFs—typical in smart homes, industrial IoT, and healthcare systems. The following sections detail the system model.}
	\subsection{Transmitter Description}\label{transmitter}
	We consider hardware imperfections at the transmitter, such as in-phase/quadrature (I/Q) modulation errors, filter distortion, and nonlinear effects. Consequently, the complex baseband transmit pilot signal can be given by $s(t) = s^I(t)+js^Q(t)$, for $t=1,\dots,J$. {Since the I and Q paths experience amplitude imbalance $\epsilon_{I/Q}$ and phase imbalance $\beta_{I/Q}$ caused by carrier modulation}, the output of the I/Q modulator can be formulated as 
	\begin{equation}\label{xpt}
		\begin{aligned}
			x(t) &= x^I(t)\cos\omega_ct- x^Q(t)\sin\omega_ct,
		\end{aligned}
	\end{equation}
	where $\omega_c$ is the carrier frequency, and we have
	\begin{equation}
		{\begin{aligned}
				x^I(t) &= s^I(t)(1 + \epsilon_I) \cos\beta_I - s^Q(t)(1 + \epsilon_Q) \sin \beta_Q,\\
				x^Q(t) &= s^I(t)(1 + \epsilon_I) \sin\beta_I + s^Q(t)(1 + \epsilon_Q) \cos\beta_Q.
		\end{aligned}}
	\end{equation}
	Upon considering the narrowband amplifier effect and utilizing a Taylor series model to represent the power amplifier, the resulting power amplifier output signal can be approximated up to the $L$-th odd order as{\cite{PA_mdl,L3}}
	\setcounter{equation}{6}
	\begin{figure*}[hb] 
		\centering 
		\vspace*{8pt} 
		\hrulefill 
		\begin{equation}
			\label{formula: alphaab}
			\alpha_{a,b}=\left\lbrace \begin{matrix}
				\sum_{k=0}^{b}\mathrm{C}_m^k\mu^{m-k}v^k\mathrm{C}_m^{b-k}(v^*)^{m+k-b}(\mu^*)^{b-k},b\leq m\\
				\sum_{k=0}^{a}\mathrm{C}_m^{m-k}\mu^kv^{m-k}s_{k,m-k}\mathrm{C}_m^{m+k-a}(v^*)^{a-k}(\mu^*)^{m+k-a}s_{a-k,m+k-a}, b \geq m
			\end{matrix}\right.		
		\end{equation}
	\end{figure*}
	\setcounter{equation}{2}
	\begin{equation}\label{model_amp}
		\overline{x}(t) \approx \sum_{l=1}^{L}\lambda_l \left[x(t)\right] ^l,
	\end{equation}
	where $\lambda_l$ is the amplifier's model parameter.
	Referring to \cite{PA_mdl} and substituting \eqref{xpt} into \eqref{model_amp}, we find that the complex envelope of \eqref{model_amp} is 
	\begin{equation}\label{model_amp1}
		\begin{aligned}
			&\tilde{x}(t)=s(t) \sum_{m=0}^{L_A}\frac{\lambda_{2m+1}}{2^{2m}}
			\mathrm{C}_{2m+1}^{m+1}\left| s(t)\right| ^{2m},
		\end{aligned}
	\end{equation}
	where $L_A=(L-1)/2$. For brevity, the subsequent derivations will use $\tilde{x}(t)$ instead of $\overline{x}(t)$. Denote $\tilde{\bm  x}=[\tilde{x}(1),\tilde{x}(2),\dots,\tilde{x}(T_s)]^T\in \mathbb{C}^{T_s}$, hence, $\tilde{\bm x}$ can be written based on \eqref{model_amp1} as
	\begin{equation}\label{mm}
		\begin{aligned}
			\tilde{\bm  x}=\widetilde{\bm S}\bm z,
		\end{aligned}
	\end{equation}
	where $\widetilde{\bm S}  = [\bm S_{L_A},\dots,\bm S_0]\in \mathbb{\bm C}^{J\times L_p}$ constructed by $s(t)$, $\bm S_m = \left[ \bm s^m_1,\dots, \bm s^m_{J}\right]^T\in \mathbb{\bm C}^{J\times (2m+2)}$ and $\bm s^m_t =  [s_t^{2m+1,0},s_t^{2m,1},\dots, s_t^{0,2m+1}]^T\in \mathbb{\bm C}^{2m+2}$ for $m = 0,1,\dots,L_A$ and $s^{a,b}_t = s(t)^a(s^*(t))^b$. Moreover, $\bm z = \bm \Delta\bm \lambda \in \mathbb{C}^{L_p}$ is the hardware feature vector, where $\bm \Delta = \text{blkdiag}\left(\overline{\bm \alpha}_{L_A},\dots,\overline{\bm \alpha}_0\right)\in \mathbb{\bm C}^{ L_p\times \left( L_A+1\right) }$, $\overline{\bm \alpha}_m = \left[\overline{\alpha}_{2m+1,0},\overline{\alpha}_{2m,1},\dots,\overline{\alpha}_{0,2m+1}\right]^T\in \mathbb{\bm C}^{2m+2}$ with
	\begin{equation}
		\overline{\alpha}_{a,b}=\left\lbrace \begin{matrix}
			\mu\alpha_{2m,0}&a=2m+1,\\
			\mu\alpha_{a-1,b}+v\alpha_{a,b-1}&0<a<2m+1,\\
			v\alpha_{0,2m}&a=0,
		\end{matrix}\right.
	\end{equation}
	where $\mu = \left((1 + \epsilon_I) e^{j \beta_I} + (1 + \epsilon_Q) e^{j \beta_Q}\right)/2, v =  \left((1 + \epsilon_I) e^{-j \beta_I} - (1 + \epsilon_Q) e^{-j \beta_Q}\right)/2$, $\alpha_{a,b}$ is defined in \eqref{formula: alphaab} at the bottom of this page; $\bm \lambda=\left[\frac{\mathrm{C}_L^{L_A+1}}{2^{2L_A}}\lambda_L,\dots,\frac{\mathrm{C}_3^2}{4}\lambda_3,\mathrm{C}_1^1\lambda_1\right]^T\in  \mathbb{\bm R}^{ L_A+1} $
	and $ L_p=(L+1)(L+3)/4$. 
	\vspace{0em}
	\subsection{Received Signal}
	We consider the {co-channel narrowband signals emitted by $K$ far-field sources that impinge on an AP with $Q$ omnidirectional antennas. Due to reflections on obstacles, the $k$-th source signal $\tilde{x}_k(t)$ has $l_k$ coherent multipath signals arriving at the antennas, where the corresponding direction of arrival (DoA) vector is $\bm \theta_k = \left[\theta_{k,1},\theta_{k,2},\dots,\theta_{k,l_k}\right]^T$, while the unknown block fading parameter is $\bm \gamma_k =[\gamma_{k,1},\gamma_{k,2},\ldots,\gamma_{k,l_k}]^T$, so that the total number of signals or DoAs is $\widetilde{K}
		= \sum_{k=1}^K{l_k}$.}
	The received signal at the AP is given as \cite{SMUSIC}
	\setcounter{equation}{7}
	\begin{equation}
		{\begin{aligned}
				\bm r(t) &= \sum_{k=1}^K\bm A_k \text{diag}(\bm \gamma_k)\bm 1_{l_k}\tilde{x}_{k}(t)\\
				&= \bm A \text{diag}(\bm \gamma) \bm \Psi \bm g(t) +\bm n(t),\quad t=1,2,\dots,J,
		\end{aligned}}
	\end{equation}
	where {$\bm A = \left[ \bm A_1, \bm A_2,\ldots,\bm A_K\right]\in\mathbb{C}^{Q\times \widetilde{K}
		}$ and $\bm A_k =\left[\bm a_{\theta_{k,1}}, \bm a_{\theta_{k,2}},\ldots,\bm a_{\theta_{k,l_k}}\right]\in\mathbb{C}^{Q\times l_k}$} is the steering matrix with $\bm a_{\theta_{k,l}} = \left[ 1,\exp(-j\phi_{1,k,l}),\ldots, \exp(-j\phi_{Q-1,k,l})\right]^T$ and $\phi_{q,k,l}=2\pi d_q \sin \theta_{k,l}/\lambda$, for $ q=1,2,\ldots,Q-1$, $k = 1,2,\ldots,K$ and $l=1,2,\ldots,l_k$, where $d_q$ is the distance from the $(q+1)$-th antenna to the reference antenna while $\lambda$ represents the wavelength, {$\widetilde{\bm \gamma} = \left[ \bm \gamma_1^T,\bm \gamma_2^T,\dots,\bm \gamma_{K}^T\right]^T$ is the block fading vector, $\bm \Psi = \text{blkdiag}(\bm 1_{l_1}, \bm 1_{l_2}, \ldots, \bm 1_{l_K})\in\mathbb{C}^{\widetilde{K}
			\times K}$}, ${\bm g} (t)=[\tilde{x} _1(t),\tilde{x} _2(t),\dots,\tilde{x} _K(t)]^T$ is the transmitted waveform vector from $K$ devices at time $t$. Collecting the received signals of $J$ snapshots in $\bm R=[\bm r(1),\dots,\bm r(J)]\in \mathbb{C}^{Q\times J}$ yields
	\begin{equation}\label{mdl}
		{\bm R=\bm A\text{diag}\left( \widetilde{\bm \gamma}\right) \bm \Psi\bm G +\bm N},
	\end{equation}
	where $\bm G= \left[\bm g (1), \bm g (2),\dots, \bm g (J) \right]= \left[\tilde{\bm  x}_1, \tilde{\bm  x}_2,\dots, \tilde{\bm  x}_K \right]^T\in \mathbb{C}^{K\times J} $ and $\bm N = \left[\bm n(1),\bm n(2),\dots,\bm n(J)\right]^T\in \mathbb{C}^{Q\times J}$. Based on the hardware model \eqref{mm} derived in Section \ref{transmitter}, we have $\tilde{\bm  x}_k =\widetilde{\bm S}_k\bm z_k$ in which the subscript $k$ refers to the $k$-th device. By substituting \eqref{mm} into \eqref{mdl}, then \eqref{mdl} can be rewritten as
	\begin{equation}\label{mdl3}
		{	\begin{aligned}
				\bm R&=\bm A\text{diag}\left( \widetilde{\bm \gamma}\right)\bm \Psi \bm Z\bm Y+\bm N,
		\end{aligned}}
	\end{equation}
	where we have $\bm Z =  \text{blkdiag} \left( \bm z_1^T,\dots,\bm z_K^T\right)  \in \mathbb{C}^{K\times KL_p}$ and $\bm Y  = \left[ 
	\widetilde{\bm S}_1,\dots,\widetilde{\bm S}_K\right]^T\in \mathbb{C}^{KL_p\times J}$. {This constitutes the generalized channel model we propose, consisting of three components: the transmitter ($\bm Z$), the propagation environment ($\text{diag}\left( \widetilde{\bm \gamma}\right)$), and the receiver ($\bm A$)}.  
	\vspace{0em}
	\section{Tensor Model-based Generalized Channel Estimation}
	{In this section, we exploit the PARAFAC decomposition and propose an iterative TALS channel estimation algorithm. Based on the unfolded forms of the received signal, the above-mentioned channel components can be estimated iteratively based on the matrix slices}. 
	\vspace{-1em}
	\subsection{Tensor Signal Modeling}
	We use \( M \) data blocks, where the \( m \)-th block is expressed as {\( \bm R_m^T = \bm V \bm \Psi^T \text{diag}\left( \widetilde{\bm \gamma}_m \right) \bm A^T + \bm N_m^T \)} for \( m = 1, \dots, M \), with \( \widetilde{\bm \gamma}_m \) denoting \( \widetilde{\bm \gamma} \) for the \( m \)-th block, and \( \bm V^T = \bm Z \bm Y \). 
	During \( M \) blocks, the incident angles remain fixed, and known pilot signals are used for \( \bm Y \). {We utilize the method described in \cite{Psi} to estimate the number of coherent multipath signals for each device $\left\lbrace l_k\right\rbrace _{k=1}^K$, based on which $\bm \Psi$ is constructed}.
	
	By stacking \( M \) matrices, we construct the three-way tensor \( \mathcal{R} \in \mathbb{C}^{J \times Q \times M} \), which can be PARAFAC decomposed as in \cite{eg2}
	\begin{equation}\label{tensor}
		{	\mathcal{R} = \mathcal{I}_{3,\widetilde{K}
			}\times_1 \left( \bm V \bm \Psi^T\right) \times_2\bm A\times_3\bm \Gamma +\mathcal{N}},
	\end{equation}
	where $\mathcal{I}_{3,\widetilde{K}}$ is a 3-way identity tensor, each dimension of which is $\widetilde{K}$, $\bm \Gamma = \left[\widetilde{\bm \gamma}_1,\dots,\widetilde{\bm \gamma}_M\right]^T\in \mathbb{C}^{M\times \widetilde{K}} $, $\mathcal{N}\in\mathbb{C}^{J\times Q\times M}$ denotes the additive Gaussian white noise tensor, and \eqref{tensor} includes the horizontal expansion matrices
	\begin{subequations}
		{
			\begin{align}
				{\bm W_1} &= \bm V \bm \Psi^T( \bm \Gamma\odot\bm A )^T+\bm N^1\in \mathbb{C}^{J\times MQ},\label{bals0}\\
				{\bm W_2} &= \bm A(\bm V \bm \Psi^T \odot \bm \Gamma)^T+\bm N^2\in \mathbb{C}^{ Q\times MJ},\label{bals1}\\
				{\bm W_3} &= \bm \Gamma(\bm A\odot \bm V \bm \Psi^T)^T+\bm N^3\in \mathbb{C}^{ M\times QJ},\label{bals2}
		\end{align}}
	\end{subequations}
	
	\noindent where $\bm N^1$, $\bm N^2$ and $\bm N^3$ are unfolded from $\mathcal{N}$.
	\vspace{-1em}
	\subsection{Alternating Least Squares Channel Estimation}
	Inspired by the PARAFAC decomposition applied in reconfigurable intelligent surfaces \cite{eg2}, we proposed an iterative TALS generalized channel estimation algorithm. The detailed steps are provided as follows:
	\subsubsection{Initialization}
	The estimated angles $\hat{\bm \theta}_0$ can be obtained via the {smoothing multiple signal classification (MUSIC) scheme\cite{SMUSIC}}, yielding the initial structured steering matrix $\hat{\bm A}_0$. Meanwhile, $\hat{\bm \Gamma}_0$ and $\hat{\bm V}_0$ {are the left singular vector matrices for the $\widetilde{K}$ non-zero singular values of $\bm W_3\bm W_3^H$ and $\bm W_1\bm W_1^H$, respectively}.
	\subsubsection{Iterative Update}
	During each iteration, the estimated generalized channels $\hat{\bm A}$, $\hat{\bm \Gamma}$ and $\hat{\bm Z}$ are obtained by the minimized structure of ALS. Starting from $\bm A$, we use the unfolded form in \eqref{bals0}. The $(i+1)$-th estimation $\hat{\bm A}_{i+1}$ is obtained from the minimization of the following regularized objective function:
	\begin{equation}\label{ls_A}
		{\begin{aligned}
				\hat{\bm A}_{i+1}&=\arg\min\limits_{\bm A}\left\|\bm W_2 -\bm A\bm B_1 \right\| ^2_F+\tau_i\left\| \bm A-\hat{\bm A}_{i}\right\| ^2_F,\\
		\end{aligned}}
	\end{equation}
	of which the iterative solution can be formulated as 
	\begin{equation}\label{so1}
		{
			\begin{aligned}
				\hat{\bm A}_{i+1}=\left(\tau_i \hat{\bm A}_{i}+\bm W_2\bm B_1^H \right)\left( \bm B_1 \bm B_1^H+\tau_i\bm I\right) ^{-1},
		\end{aligned}}
	\end{equation}
	where  $\bm B_1=\left(  \bm Y^T\hat{\bm Z}_i^T\bm \Psi^T \odot \hat{\bm \Gamma}_i \right) ^T $  and the regularization coefficient $\tau_i$ of the $i$-th iteration is reduced from an initial value $\tau_0 = 0.1$ by a factor of $\delta = 0.9$, following the relation $\tau_i = \delta \tau_{i-1}$.
	Consequently, $\hat{\bm A}_{i+1}$ can be updated by estimating $\hat{\bm \theta}_{i+1}$ from $\hat {\bm A}_{i+1}$ through beamforming. Then, we can obtain $\hat {\bm A}_{i+1}$ according to a steering matrix formulation. Similarly, we have
	\vspace{-1em}
	\begin{subequations}
		{
			\begin{align}
				\hat{\bm Z}_{i+1}& =\arg\min\limits_{\bm Z}\left\|	\bm W_1-\bm Y ^T\bm Z^T\bm B_2 \right\| ^2_F+\tau_i\left\| \bm Z-\hat{\bm Z}_{i} \right\| ^2_F,\\
				\hat{\bm \Gamma}_{i+1}&=\arg\min\limits_{\bm \Gamma}\left\|\bm W_3 - \bm \Gamma\bm B_3 \right\| ^2_F+\tau_i\left\| \bm  \Gamma-\hat{\bm \Gamma}_i\right\| ^2_F,\label{ls_gma}
		\end{align}}
	\end{subequations}
	
	\noindent the solutions of which are respectively formulated as
	\begin{subequations}
		{
			\begin{align}
				\hat{\bm z}^T_{i+1}  &= \left( \tau_i \bm I +\bm C^H\bm C \right)^{-1}  \left[ \bm C^H\mathrm{vec}\left(\bm W_1 \right)+\tau_i\mathrm{vec}\left( \bm Z_i\right)  \right]  ,\label{so2}\\
				\hat{\bm \Gamma}_{i+1}&=\left(\tau_i \hat{\bm \Gamma}_i+\bm W_3\bm B_3^H \right)\left( \bm B_3 \bm B_3^H+\tau_i\bm I\right) ^{-1}, \label{so3}
		\end{align}}
	\end{subequations}
	
	\noindent where we have $\hat{\bm z}^T_{i+1}=\mathrm{vec}\left(\hat{\bm Z}^T_{i+1}\right)$, $\bm B_2=\left( \hat{ \bm \Gamma}_i\odot\hat{\bm A}_i \bm \Psi\right) ^T$, $\bm B_3=\left( \hat{\bm A}_i \odot \bm Y^T\hat{\bm Z}_i^T\bm \Psi^T\right) ^T $ and $\bm C = \bm B_2^H \otimes \bm Y^H$.
	\subsubsection{Iteration Termination Criterion}
	The algorithm stops when either the maximum number of iterations is reached or the relative reconstruction error satisfies $\left|\text{loss}_{i+1}-\text{loss}_i\right| /\text{loss}_i<\rho$, where $\rho$ is the threshold and we have $\text{loss}_i=\left\| {\bm W_1} - {\bm A}_i( {\bm \Gamma}_i\odot{\bm V }_i)^T\right\|$. It should be noted that when solving using the ALS method, there is typically a scaling ambiguity. However, when $\epsilon_I = -\epsilon_Q = \epsilon$ and $\beta_I = -\beta_Q= \beta$ for the $k$-th target, we conveniently ensure that the second-to-last element of $\bm z_k$ equals 1, thus resolving the ambiguity and making it possible to solve the problem.
	\subsection{Computational Complexity}
{For the TALS methods discussed in Section III-B, the computational costs associated with steps \eqref{so1}, \eqref{so2}, and \eqref{so3} are $O\left(QMJ\widetilde{K}+\widetilde{K}^2MJ \right)$, $O\left(K^4L_p^2MJQ\right)$, and $O\left(MQJ\widetilde{K}+\widetilde{K}^2QJ\right)$, respectively. Specifically, the matrix inversion complexities for these steps are $O(\widetilde{K}^3)$, $O(K^6L_p^3)$, and $O(\widetilde{K}^3)$, respectively, with step \eqref{so2} being the most computationally intensive. These operations are particularly demanding for large-scale problems where \(K\), \(L_p\), or \(\widetilde{K}\) becomes significantly large. In this study, we focus on scenarios with moderate dimensions, ensuring that the computational requirements remain manageable within the available resources. While the current setup is sufficient for validating the theoretical framework, future work could explore optimization techniques, such as approximate matrix inversion or dimensionality reduction, to address the challenges associated with larger-scale systems.}
		 
	\setcounter{equation}{25}
	\begin{figure*}[ht] 
		\centering 
		\vspace*{1pt} 
		\begin{equation}\label{M}
			{
				\bm M^{a,x}_{b,y}=\sigma^2\left[  \begin{matrix}
					&\bm 0_{(y-1)\times ZX}&\\
					&\bm 1_{1\times Z}&\\
					\bm 0_{\left( (Z-1)Y+1\right) \times Z(x-1)}&\vdots&\bm 0_{\left( (Z-1)Y+1\right) \times  Z(X-x)}\\
					&\bm 1_{1\times Z}&\\
					&\bm 0_{(Y-y)\times ZX}&\\
				\end{matrix} \right]}  
		\end{equation}
		\hrulefill
		\vspace{-1em}
	\end{figure*}
	\setcounter{equation}{16}
	\subsection{CRLB of TALS Channel Estimation Method}\label{app:crlb_der}
	As derived from \eqref{bals0}, \eqref{bals1} and \eqref{bals2}, the likelihood functions can be formulated as
	\begin{equation}
		{
			\begin{aligned}
				\mathcal{L} &=\eta\exp\left\lbrace -\frac{1}{\sigma^2}\sum_{j=1}^{J}\left\| {\bm W}^1_{j,:}-\left( \bm V\bm \Psi^T \right) _{j,:}\left( \bm \Gamma \odot \bm A\right) ^T\right\|^2 \right\rbrace \\
				&=\eta\exp\left\lbrace -\frac{1}{\sigma^2}\sum_{q=1}^{Q}\left\| {\bm W}^2_{q,:}-\bm A_{q,:}\left( \bm V\bm \Psi^T \odot \bm \Gamma \right) ^T\right\|^2 \right\rbrace \\
				&=\eta\exp\left\lbrace -\frac{1}{\sigma^2}\sum_{m=1}^{M}\left\| {\bm W}^3_{m,:}-\bm \Gamma_{m,:}\left( \bm A \odot \bm V\bm \Psi^T \right) ^T\right\|^2 \right\rbrace,
		\end{aligned}}
	\end{equation}
	where $\eta = 1/(\pi\sigma^2)^{QMJ}$.
	
	To derive the CRLB of complex parameters, we introduce the following complex parameter vector $\bm \xi  \in  \mathbb{C}^{\widetilde{K}
		+2Kl^1_p+2M\widetilde{K}
	}$, yielding
	\begin{equation}
		\bm \xi \triangleq  [{\bm \theta },\underbrace{\overline{\bm z}_1^T,\dots,\overline{\bm z}_K^T,\bm \gamma_1^T,\dots,\bm \gamma_M^T}_{\bm \xi_1},\underbrace{\overline{\bm z}_1^H,\dots,\bm \gamma_M^H}_{\bm \xi^*_1}]^T,
	\end{equation}
	where $\overline{\bm z}_k = [\bm z_{k}(1),\dots,\bm z_{k}(L_p-2),\bm z_{k}(L_p)]^T\in \mathbb{C}^{L_p^1}$ with $L_p^1 = L_p-1$ to differentiate from $\bm z_k$. Moreover, we have $\bm \gamma_m = \bm \Gamma^T_{:,m}$ for $m=1,\dots,M$. 
	Taking the partial derivatives of $\ln \mathcal{L}$ with respect to the unknown complex parameters in $\bm \xi$ yields
	\begin{equation}
		{
			\begin{aligned}
				&\frac{\partial \ln \mathcal{L}}{\partial \theta_k}=-\frac{2\cos \theta_k}{\sigma^2}\sum_{q=1}^{Q} \Re \left[ \bm N^{2*}_{q,:}( \bm V_1\odot \bm \Gamma)_{:,k}\beta_q\bm A_{q,k}\right],  \\
				&\frac{\partial \ln \mathcal{L}}{\partial \bm \gamma_m(k)  }=\frac{1}{\sigma^2}\mathbf{e}_k^T (\bm A \odot \bm V_1 )^T\bm N^{3H}_{:,m},\\
				&\frac{\partial  \ln \mathcal{L}}{\partial \overline{\bm z}_k(p)}=\frac{1}{\sigma^2}\sum_{j=1}^{J}\bm \Psi^T_{:,k}\bm Y^T_{j,d}\left( \bm \Gamma \odot \bm A\right) ^T\bm N^{1H}_{:,j},\\
		\end{aligned}}
	\end{equation}
	and
	\begin{equation}
		\begin{aligned}
			\frac{\partial  \ln \mathcal{L}}{\partial  \overline{\bm z}^*_k(p)} &= \left(	\frac{\partial  \ln \mathcal{L}}{\partial  \overline{\bm z}_k(p)}  \right) ^*
			\frac{\partial \mathcal{L}}{\partial  \bm \gamma^*_m(k)}  &= \left( \frac{\partial \mathcal{L}}{\partial  \bm \gamma_m(k)}  \right) ^*,
		\end{aligned}
	\end{equation}
	where $\beta_q = {j2\pi(q-1)h}/{\lambda}$ with the antenna spacing $h$. Moreover, we have $d=(k-1)\times L_p +p$, when $p\neq L^1_p$ and $p=p+1$ in the case of $p=L^1_p$.
	
	The Fisher information matrix can be expressed as
	\begin{equation}\label{fim}
		\begin{aligned}
			\bm F &= \mathbb{E}\left[ \left(\frac{\partial \ln L }{\partial \bm \xi } \right)  \left(\frac{\partial \ln L }{\partial \bm \xi } \right) ^H\right] =\left[  \begin{matrix}
				\bm F_1  & \bm  F_3^H \\
				\bm  F_3& \bm  F_2
			\end{matrix} \right],\\
		\end{aligned}
	\end{equation}
	where we have
	\begin{equation}
		\begin{aligned}
			&\bm F_1(k_1,k_2) =  \mathbb{E} \left[  \frac{\partial \ln \mathcal{L}}{\partial \theta_{k_1}}\frac{\partial \ln \mathcal{L}}{\partial \theta_{k_2}} \right]= \frac{2\cos \theta_{k_1}\cos \theta_{k_2}}{\sigma^4}\Re\left[ \zeta \right], 
		\end{aligned}
	\end{equation}
		where { $\zeta= \sum_{q=1}^{Q}\beta_{q}\beta_{2-q}\bm A_{q,k_1}\bm A^*_{q,k_2}\left(  \bm V_1\odot \bm \Gamma\right) ^T_{k_1,:}\left(  \bm V_1\odot \bm \Gamma\right) ^*_{:,k_2}$}, 
		while $\bm F_2 \in  \mathbb{C}^{(2KL^1_p+2M\widetilde{K})\times (2KL^1_p+2M\widetilde{K})}$ in \eqref{fim} is given by 
		\begin{equation}
			\begin{aligned}
				\bm F_2 &= \left[  \begin{matrix}
					\bm M_1 & \bm M_2^* \\
					\bm M_2 & \bm M_1^* \\
				\end{matrix} \right],  
			\end{aligned}
		\end{equation}
		where $\bm M_1 \in  \mathbb{C}^{\left( KL^1_p+M\widetilde{K}
			\right) \times \left( KL^1_p+M\widetilde{K}
			\right) }$ is defined as
		\begin{equation}
			\begin{aligned}
				\bm M_1  &=E\left[  \begin{matrix}
					\frac{\partial \ln \mathcal{L}}{\partial \overline{\bm z}} \left( \frac{\partial \ln \mathcal{L}}{\partial   \overline{\bm z}}\right) ^H  & 	\frac{\partial \ln \mathcal{L}}{\partial \overline{\bm z}} \left( \frac{\partial \ln \mathcal{L}}{\partial   \bm \gamma }\right) ^H  \\
					\frac{\partial \ln \mathcal{L}}{\partial \overline{\bm \gamma}} \left( \frac{\partial \ln \mathcal{L}}{\partial   \overline{\bm z}}\right) ^H &\frac{\partial \ln \mathcal{L}}{\partial \overline{\bm \gamma }} \left( \frac{\partial \ln \mathcal{L}}{\partial   \bm \gamma }\right) ^H   \\
				\end{matrix} \right]  =  \left[  \begin{matrix}
					\bm O_1 & \bm O_2^H\\
					\bm O_2 & \bm O_3 \\
				\end{matrix} \right],  
			\end{aligned}
		\end{equation}
		where $\bm O_1 \in  \mathbb{C}^{(KL^1_p)\times (KL^1_p)} $, $\bm O_2 \in  \mathbb{C}^{M\widetilde{K}\times (KL^1_p)} $ and  $\bm O_3 \in  \mathbb{C}^{M\widetilde{K}\times M\widetilde{K}} $ are given by
		\begin{equation}\label{o1}
			{
				\begin{aligned}
					&\bm O_1\left( (k_1-1)L^1_p+p_1,(k_2-1)L^1_p+p_2\right) \\
					&= \left\lbrace \begin{array}{l}
						\frac{1}{\sigma^2}\bm{\Psi}^T_{:,k}\sum_{j=1}^{J}\bm{Y}^T_{j,d_1}(\bm{\Gamma} \odot \bm{A})^T (\bm{\Gamma} \odot \bm{A})^*\bm{Y}^H_{j,d_2}\bm{\Psi}_{k,:},\\
						0,\text{else},
					\end{array} \right.\\
					&\bm O_2\left( (m-1)\widetilde{K}+k_1,(k_2-1)L^1_p+p\right) \\
					& =  \frac{1}{\sigma^2} \mathbf{e}_{k_1}^T (\bm{A} \odot \bm{V}_1 )^T\sum_{j=1}^{J}\bm{M}^{3,m}_{1,j}(\bm{\Gamma} \odot \bm{A})^*\bm{Y}^H_{j,d}\bm{\Psi}_{k_2,:},\\
					&\bm O_3\left( (m_1-1)\widetilde{K}+k_1,(m_2-1)\widetilde{K}+k_2\right) \\
					&= \left\lbrace \begin{array}{l}
						\frac{1}{\sigma^2}(\bm{A} \odot \bm{V}_1 )^T_{k_1,:} (\bm{A} \odot \bm{V}_1 )^*_{:,k_2},\\
						0,m_1 \neq m_2.
					\end{array} \right.
			\end{aligned}}
		\end{equation}
		{Define $\bm M^{1,j}_{2,q} = \mathbb{E}\left[  (\bm N^{1}_{:,j})^H \bm N^2_{q,:} \right] \in  \mathbb{C}^{MQ \times MJ}$, $\bm M^{2,q}_{3,m} =\mathbb{E}\left[ (\bm N^{2}_{:,q})^H  \bm N^3_{m,:}\right] \in  \mathbb{C}^{MJ\times QJ } $ and $\bm M^{3,m}_{1,j} = \mathbb{E}\left[ (\bm N^{3}_{:,m})^H  \bm N^1_{j,:} \right]  \in  \mathbb{C}^{QJ\times MQ}$} in \eqref{M} at the top of this page, where $a\in\{1,2,3\}$ and $X\in\{Q,J,M\}$ for $x=1,\ldots,X$, and it is the same for all values of $b$ and $y$. Here, $Z$  is the remaining element in $\{Q, J, M\}$ after excluding the components corresponding to $X$ and $Y$. Also, $\bm M^*_2 \in  \mathbb{C}^{(Kl^1_p+M\widetilde{K}
			)\times (Kl^1_p+M\widetilde{K}
			)}$ is defined as
		\setcounter{equation}{26}
		\begin{equation}
			{
				\begin{aligned}
					&\bm M_2^* =\mathbb{E}\left[  \begin{matrix}
						\frac{\partial \ln \mathcal{L}}{\partial \overline{\bm z}} \left( \frac{\partial \ln \mathcal{L}}{\partial   \overline{\bm z}}\right) ^T  & 	\frac{\partial \ln \mathcal{L}}{\partial \overline{\bm z}} \left( \frac{\partial \ln \mathcal{L}}{\partial   \bm \gamma }\right) ^T  \\
						\frac{\partial \ln \mathcal{L}}{\partial {\bm \gamma}} \left( \frac{\partial \ln \mathcal{L}}{\partial   \overline{\bm z}}\right) ^T &\frac{\partial \ln \mathcal{L}}{\partial {\bm \gamma }} \left( \frac{\partial \ln \mathcal{L}}{\partial   \bm \gamma }\right) ^T   \\
					\end{matrix} \right],  \\
			\end{aligned}}
		\end{equation}
		composed of expectations 
		all equal to zero\cite{SPF}, implying that $\bm M_2^* =\bm M_2= \bm 0$. Consequently, $\bm F_3 \in  \mathbb{C}^{(2KL^1_p+2M\widetilde{K}
			)\times \widetilde{K}
		}$ in \eqref{fim} can be formulated as 
		\begin{equation}\label{f3}
			{
				\begin{aligned}
					\bm F_3 & =\mathbb{E}\left[    \left[  \begin{matrix}
						\frac{\partial \ln \mathcal{L}}{\partial \bm \xi_1}  \\
						\frac{\partial \ln \mathcal{L}}{\partial \bm \xi_1^*}\\
					\end{matrix} \right]   \left( \frac{\partial \ln \mathcal{L}}{\partial  \bm \theta}\right) ^H  \right] \\ 
					&= \left[  \begin{matrix}
						(\bm M_3^{1})^T & (\bm M_3^{2})^T & (\bm M_3^{1})^H & (\bm M_3^{2})^H
					\end{matrix} \right]^T,  
			\end{aligned}}
		\end{equation}
		where $\bm M_3^1 \in \mathbb{C}^{Kl_p^1\times \widetilde{K}}$ and $\bm M_3^2 \in \mathbb{C}^{M\widetilde{K}\times \widetilde{K}}$ are respectively defined as
		\begin{subequations}
			{
				\begin{align}
					\bm M_3^1((k_1-1)L^1_p+p,k_2)&=
					-\frac{\cos \theta_{k_2}}{\sigma^4}\bm \Psi^T_{:,k_1}\kappa ,\\
					\bm M_3^2((m-1)\widetilde{K}+k_1,k_2)
					&=-\frac{\cos \theta_{k_2}}{\sigma^4}\mathbf{e}_{k_1}^T(\bm A \odot \bm V_1 )^T\chi,
			\end{align}}
		\end{subequations}
		where
		\begin{subequations}
			{
				\begin{align}
					\kappa &=\sum_{j=1}^{J}\sum_{q=1}^{Q}\bm Y^T_{j,idx}(\bm \Gamma \odot \bm A)^T \bm M^{1,j}_{2,q}\beta_{2-q}\bm A^*_{q,k_2}( \bm V_1\odot \bm \Gamma)^*_{:,k_2},\\
					\chi &= \sum_{q=1}^{Q}\left( \bm M^{2,q}_{3,m}\right) ^H\beta_{2-q}\bm A^*_{q,k_2} ( \bm V_1\odot \bm \Gamma)^*_{:,k_2}.
			\end{align}}
		\end{subequations}
		
		By using the formula for the inverse of a block matrix \cite{matbook}, the targeted CRLBs are
		\begin{subequations}
			\begin{align}
				\textbf {CRLB}(\bm \theta)&=\bm  F_1^{-1}+\bm  F_1^{-1}\bm  F_3^H \bm C_1\bm  F_3\bm  F_1^{-1},\label{crlb_theta}\\
				\textbf {CRLB}(\bm \xi,\bm \xi^*) & =\bm  F_2^{-1}+\bm F_2^{-1}\bm  F_3\bm C_2\bm  F_3^H\bm F_2^{-1},\label{crlb_xi}
			\end{align}
		\end{subequations}
		where $\textbf {CRLB}(\cdot)$ is the CRLB for parameters $\bm C_1 = \bm  F_1-\bm  F_3^H\bm F_2^{-1}\bm  F_3$ and $\bm C_2 = \bm  F_2-\bm  F_3\bm F_1^{-1}\bm  F_3^H$.
		\vspace{-1em}
		\section{Simulation Results}
		Considering \( K = 3 \) devices, their transmitted signals arrive at the AP through {\( l_1 = 1 \), \( l_2 = 2 \), and \( l_3 = 2 \) distinct paths, with corresponding DoAs \( \bm \theta_1 = [-24.82^\circ]^T \), \( \bm \theta_2 = [-3.57^\circ, 17.96^\circ]^T \), and \( \bm \theta_3 = [25.72^\circ, 40.81^\circ]^T \).} The block fading parameters \( \bm \gamma_1, \bm \gamma_2, \bm \gamma_3 \) depend on the SNR for their modulation schemes, with phases uniformly distributed in \([-\pi, \pi]\). The I/Q gains are \( \epsilon = [0.0001, -0.0028, -0.0051] \) corresponding to {percentage values of \([0.01\%, -0.28\%, -0.51\%]\)} and phase errors \( \beta = [-0.018^\circ, 0.0175^\circ, 0.0120^\circ] \). The power amplifier order is \( L = 3 \), with parameter matrix $\bm \Lambda=[1,0,0.3;1,0,0.6;1,0,0.4]$, where each row corresponds to a device. Baseband signals adopt QPSK modulation at carrier frequency {\( f_c = 2.4 \, \text{GHz} \)}. The noise \( \bm N \) in \eqref{mdl} has zero mean and unit variance elements. An {eight-element} ULA, with half-wavelength spacing, is deployed at the receiver. The iteration control threshold is set to \( \rho = 10^{-10} \). We compare the Root Mean Square Error (RMSE) of the proposed TALS method with both {KRF in \cite{eg2}} and LS. For DoA estimation, we define {$\text{RMSE}(\bm \theta) = \sqrt{\frac{1}{P}\sum_{p=1}^{P}\sum_{k=1}^{K} \left\|\hat{\bm \theta}_{k,p}-\bm\theta_k\right\|^2}$}, where $P$ denotes the number of Monte-Carlo simulations, while $\hat{\bm\theta}_{k,p}$ is the estimate of $\bm \theta_k$ in the $p$-th trial. The $\text{RMSE}(\bm z)$ is defined similarly. The CRLBs given by \eqref{crlb_theta} and \eqref{crlb_xi} are also plotted for comparison. 
		\begin{figure}[htbp]
			\vspace{0em}
			\centering
			\includegraphics[width=0.38\textwidth]{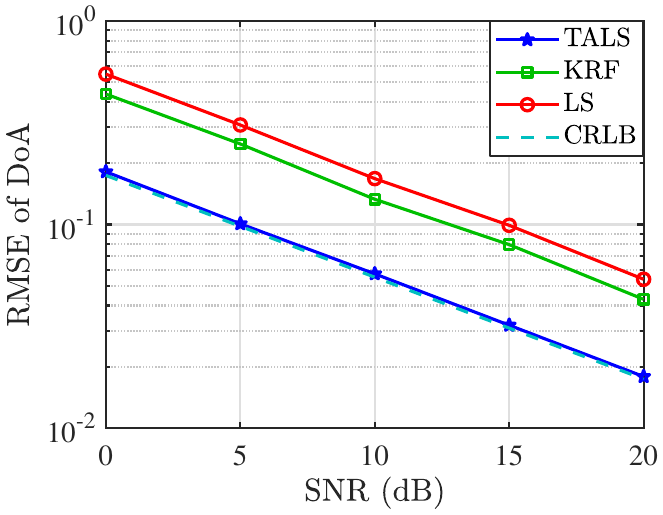} 
			\caption{RMSE of DoA versus the SNR.}
			\label{fig1}
			\vspace{-2em}
		\end{figure}    
		
		Figure \ref{fig1} shows the RMSE performance for DoA estimation using the proposed TALS, {KRF}, LS, and the CRLB as the theoretical lower bound. The results indicate that TALS consistently achieves superior performance compared to KRF and LS, closely approaching the CRLB at all SNR levels. At an RMSE of $10^{-1}$, {TALS attains an SNR gain of about 10 dB over LS and 7.8 dB over KRF}, highlighting its robustness and efficiency. This advantage stems from TALS’s capacity to exploit the underlying signal structure for precise DoA estimation—a key feature for high-precision applications. By contrast, KRF and LS deviate notably from the CRLB, underscoring their limitations under similar conditions.
		\begin{figure}[htbp]
			\vspace{0em}
			\centering
			\includegraphics[width=0.38\textwidth]{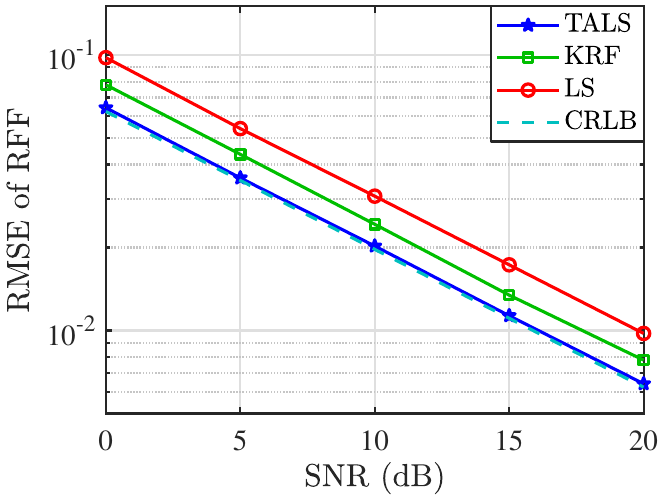}
			\caption{RMSE of RFF versus the SNR.}
			\label{fig2}
			\vspace{-1em}
		\end{figure}
		
		Figure \ref{fig2} presents the RMSE performance for RFF ($\bm Z$) estimation. As with the DoA results, TALS significantly outperforms LS and {KRF} across all SNR levels. Furthermore, as SNR increases, TALS almost coincides with the CRLB, highlighting its near-optimal performance. This robustness and precision make TALS a reliable method for RFF feature extraction, which is crucial for accurate device identification. These results demonstrate TALS’s potential for real-world deployment in applications requiring extreme precision and noise robustness.
		
		\begin{figure}[htbp]
			\vspace{0em}
			\centering
			\includegraphics[width=0.38\textwidth]{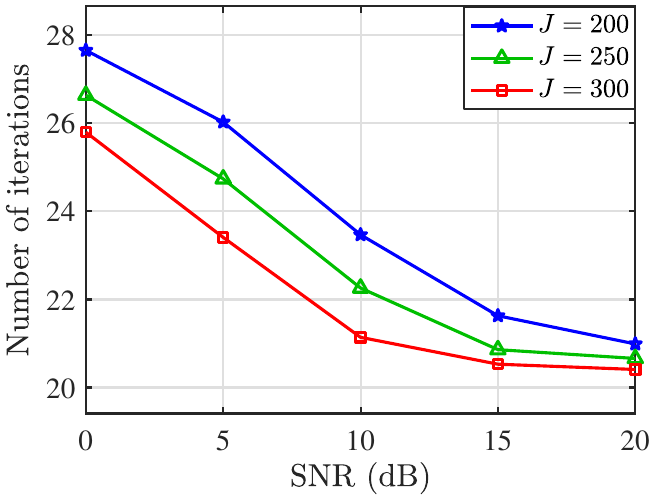}
			\caption{Convergence of the proposed TALS algorithm.}
			\label{fi3}
			\vspace{-1em}
		\end{figure}     
		In Fig. \ref{fi3}, we evaluate the number of TALS iterations according to the criterion discussed in Section III-B. This result verifies that our algorithm requires only a few iterations to stop.
		
\begin{figure}[htbp]
			\vspace{0em}
			\centering
			\includegraphics[width=0.40\textwidth]{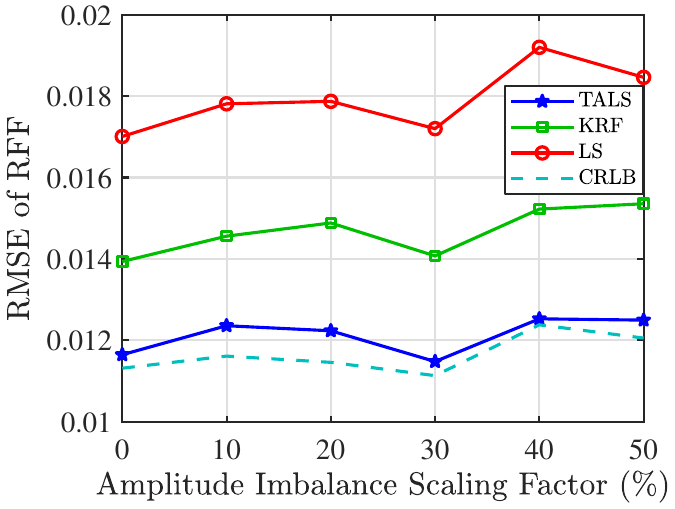}
			\caption{RMSE of RFF versus the amplitude imbalance scaling factor.}
			\label{fi4}
			\vspace{-1em}
		\end{figure}

\begin{figure}[htbp]
			\vspace{-1em}
			\centering
			\includegraphics[width=0.40\textwidth]{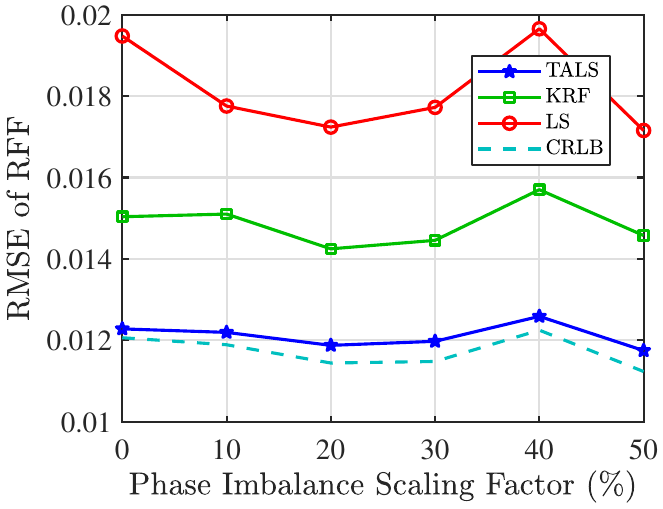}
			\caption{RMSE of RFF versus the phase imbalance scaling factor.}
			\label{fi5}
			\vspace{-1em}
		\end{figure}

		{Figures \ref{fi4} and \ref{fi5} illustrate the relationship between the RMSE and the percentage of amplitude and phase imbalance, respectively. The RMSE exhibits slight fluctuations with increased imbalance scaling factors, although the overall RMSE remains minimal. The above trends demonstrate the robustness of the estimation process, as the performance is not significantly affected by variations in amplitude or phase imbalance. The minor fluctuations in the RMSE are within acceptable margins and do not impact the overall accuracy of the estimation.}
		
		\section{Conclusion}
		A novel multiple device identification model under multipath propagation was conceived. By exploiting the PARAFAC decomposition of the received tensor-based signals, we developed an iterative TALS channel estimation algorithm to recover the DoA, channel fading, and hardware-related parameters jointly. We also derived the corresponding CRLB as a theoretical performance benchmark. Our simulation results demonstrated that the proposed algorithm outperforms systematically both the {KRF} and LS methods with good convergence.
		\bibliographystyle{bibtex/IEEEtran}	
\bibliography{IEEEabrv,main}
	\end{document}